\documentclass[11pt]{article}

\usepackage[margin=1in]{geometry}
\usepackage{amsmath,amssymb,amsfonts}
\usepackage{graphicx}
\usepackage{booktabs}
\usepackage{natbib}
\usepackage{microtype}
\usepackage{xcolor}
\usepackage[hidelinks]{hyperref}
\usepackage{url}

\newcommand{\Kn}{\mathit{Kn}}
\newcommand{\dd}{\mathrm{d}}
\newcommand{\cl}{\mathrm{cl}}
\newcommand{\DVM}{\mathrm{DVM}}

\newcommand{\tr}{\operatorname{tr}}

\title{Closure-channel identifiability and two-channel recovery in monatomic kinetic normal shocks}
\author{Ehsan Roohi\\
Department of Mechanical and Industrial Engineering\\
University of Massachusetts Amherst\\
Amherst, MA 01003, USA\\
\texttt{eroohi@umass.edu}}
\date{}

\begin{document}
\maketitle

\begin{abstract}
Residual agreement in a kinetic or moment equation does not automatically identify every higher-order closure variable entering a nonequilibrium shock.  We formulate this issue as an observability problem for the fourth-order closure content of monatomic normal shocks and follow it through a hierarchy of collision models and diagnostics.  The kinematic part of the result is independent of the collision operator: the one-dimensional heat-flux budget observes the projected fourth-order channel $S=R^{\cl}_{xx}+\Delta/3$, not the tensorial R26-level moment $R^{\cl}_{xx}$ separately from the scalar fourth-order excess $\Delta$.  The observation map therefore has a one-dimensional null space, so a heat-flux residual can be small while the split between tensorial anisotropy and isotropic tail intensity remains wrong.  A DVM-consistent scalar-excess budget supplies the missing channel and gives the two-channel reconstruction $R^{\cl}_{xx}=S-\Delta/3$ without direct $R^{\cl}_{xx}$ data.  Across BGK shocks at Mach 2--5, this reduces the active-zone $R^{\cl}_{xx}$ error from about $63$--$64\%$ to $2.4$--$4.1\%$.  Sparse scalar-excess interpolation is used only as an information-reduction test: a representative 24-probe operating point gives $R^{\cl}_{xx}$ errors below $4.5\%$, and below $4.7\%$ with $1\%$ probe noise.  Collision-model diagnostics then separate the invariant observation channel from the model-dependent source law.  Shakhov changes the heat-flux relaxation to the correct Prandtl number but is neutral in the even $|\boldsymbol c|^4$ scalar-excess source; a direct discrete Shakhov channel check recovers $S$, $\Delta$ and $R^{\cl}_{xx}$ with errors $6.4\times10^{-4}$, $2.1\times10^{-7}$ and $1.0\times10^{-3}$, respectively.  ES-BGK adds a stress-quadratic scalar-excess target contribution of $4.3$--$4.5\%$ of $\|\Delta\|$, whereas the ES-FP anisotropic diffusion source gives a larger correction of about $21.5$--$22.4\%$ while remaining strongly correlated with the BGK/Shakhov source.  Finally, an eight-realisation DSMC Mach-3 ensemble accumulates $|\boldsymbol c|^4$ production directly from accepted binary collisions; the collisional production closes the scalar-excess spatial budget with correlation $0.987$ and is strongly correlated with $-\Delta$ ($\rho=0.968$).  The result is a closure-identifiability principle: the heat-flux equation fixes a fourth-order energy-transport channel, while the tensorial and scalar closure variables require an independent scalar complement whose source law depends on the collision model.
\end{abstract}

\section{Introduction}
\label{sec:introduction}

A rarefied shock is a velocity-space object.  Its density, velocity and temperature are low-order projections of the molecular distribution, while closure variables in Grad-type and regularised moment systems are higher-order, sign-changing and tail-weighted projections.  The latter are especially sensitive to high-peculiar-speed tails, where sign-changing cancellations are almost invisible to density and velocity but dominate fourth-order transport.  This distinction is central in kinetic gas theory \citep{Cercignani1988,Sone2007,Bird1994} and in reduced moment models descended from Grad's expansion \citep{Grad1949,Levermore1996,Struchtrup2005,Torrilhon2016}.  In the regularised 13- and 26-moment hierarchies, higher moments are not optional diagnostics: the stress equation transports third-order information, and the heat-flux equation transports fourth-order information \citep{StruchtrupTorrilhon2003,TorrilhonStruchtrup2004,StruchtrupTorrilhon2007PRL,GuEmerson2009R26}.  In R26, the third-order tensor $m_{ijk}$, the fourth-order tensor $R_{ij}$ and the scalar fourth-order excess $\Delta$ are promoted to closure-level variables.  A kinetic solver or reduced model that reproduces the visible shock profile but misses these variables has not recovered the closure information needed by the moment hierarchy.

Normal shocks are a particularly stringent test of this issue.  They have long served as kinetic benchmarks for the Boltzmann equation, DSMC, moment closures and model kinetic equations because the shock layer concentrates transport, production and tail populations into a narrow spatial region \citep{MottSmith1951Shock,Alsmeyer1976Shock,IvanovGimelshein1998Hypersonic}.  Recent comparisons of BGK, Shakhov, ES-BGK and Fokker--Planck variants confirm that models with similar lower-order transport can differ substantially in higher-order moments in shock waves \citep{FeiLiuLiuZhang2020ShockBenchmark}.  This makes the normal shock an appropriate base problem for asking not only whether a solver reproduces low-order fields, but also which closure content its residuals actually identify.

This issue is increasingly relevant because residual-based and physics-informed neural solvers are now used for kinetic equations \citep{Raissi2019PINN,Karniadakis2021PINNs,LouMengKarniadakis2021PINNBGK,Oh2025SPINNBGK}.  Such solvers are commonly validated using low-order profiles or small residuals in selected moment equations.  That practice is potentially misleading in rarefied flows: a residual can be exactly satisfied while observing only a projected part of the nonequilibrium distribution.  A companion neural-shock study \citep{Roohi2026NeuralShock} showed this failure mode empirically: a residual-consistent BGK shock solver could recover the macroscopic profiles and several lower nonequilibrium moments while leaving the tensorial fourth-order closure moment poorly determined.  The present paper deliberately removes the neural architecture from the centre of the argument.  Its purpose is to identify the information content of the kinetic budgets themselves.  The question addressed here is therefore not whether a residual-based kinetic model can fit a smooth shock profile, but which fourth-order closure content is actually observable from the heat-flux budget of the shock.  The answer is structural: the heat-flux budget selects a combined channel, while recovery of the tensorial closure moment itself requires one additional scalar fourth-order channel.

The present paper is therefore a theoretical and computational identifiability study, rather than a new neural architecture.  Stationary monatomic normal shocks provide a controlled setting in which full discrete-velocity-method (DVM) distributions, moment budgets and collision-model variants can be evaluated without the ambiguity of sampling noise.  This setting also separates two questions that are often conflated.  The first is kinematic: which fourth-order projection is present in the heat-flux flux?  The second is collisional: how does a chosen kinetic model produce or relax the missing scalar-excess channel?  The first question is independent of the collision operator; the second distinguishes a hierarchy of models, from BGK relaxation, through correct-Prandtl Shakhov, ES-BGK and ES-FP targets, to direct DSMC sampling of Boltzmann-like binary collisions.  That hierarchy is important for the present paper.  BGK is the cleanest setting for deriving and validating the two-budget mechanism; Shakhov tests whether the conclusion survives correction of the heat-flux relaxation; ES-BGK tests whether anisotropic covariance changes the scalar-excess source; ES-FP tests whether an anisotropic diffusion process modifies the same scalar channel more strongly; and DSMC tests whether the scalar-excess production can be measured directly from molecular collisions rather than imposed by any relaxation or Fokker--Planck ansatz.

We first establish the heat-flux observation operator and its null space.  We then derive the complementary scalar-excess budget, verify the two-budget recovery across Mach number, test sparse and noisy scalar-excess information, and check collision-model dependence using Shakhov, ES-BGK and ES-FP source diagnostics.  Finally, we introduce a standalone DSMC shock calculation in which the change of $|\boldsymbol c|^4$ is accumulated over accepted binary collisions.  The contribution is not a new R26 balance law and not a second demonstration of a neural solver.  It is a closure-identification result: a rank-deficient fourth-order channel is converted into a constructive two-channel recovery principle, and the missing scalar channel is traced from model relaxation sources to direct collision production.

\section{Kinetic shocks, R26 moments and DVM references}
\label{sec:reference}

We begin with the steady one-dimensional BGK model \citep{Bhatnagar1954},
\begin{equation}
  v_x\partial_x f = \frac{1}{\Kn}\{M[f]-f\},
  \label{eq:bgk}
\end{equation}
where $f(x,\boldsymbol v)$ is the distribution and $M[f]$ is the local Maxwellian with the same density $\rho$, streamwise velocity $u_x$ and temperature $T$ as $f$.  The peculiar velocity is $\boldsymbol c=\boldsymbol v-u_x\boldsymbol e_x$.  The heat flux and normal stress are
\begin{equation}
  q_x = \frac{1}{2}\int |\boldsymbol c|^2 c_x f\,\dd\boldsymbol v,
  \qquad
  \sigma_{xx}=\int \left(c_x^2-\frac{|\boldsymbol c|^2}{3}\right)f\,\dd\boldsymbol v .
  \label{eq:q_sigma}
\end{equation}
The fourth-order quantities of interest are
\begin{equation}
  R^{\cl}_{xx}
  =\int |\boldsymbol c|^2\left(c_x^2-\frac{|\boldsymbol c|^2}{3}\right)(f-M_d)\,\dd\boldsymbol v
   - 7T\sigma_{xx},
  \label{eq:Rdef}
\end{equation}
with $M_d$ the conservative discrete Maxwellian on the same velocity grid, and
\begin{equation}
  \Delta_d=\int |\boldsymbol c|^4(f-M_d)\,\dd\boldsymbol v .
  \label{eq:Deltadef}
\end{equation}
This is the DVM-consistent counterpart of $\int |\boldsymbol c|^4 f\,\dd\boldsymbol v-15\rho T^2$.  In a sufficiently resolved velocity grid the two differ only by the equilibrium quadrature offset.  We remove the plateau offset consistently from both $R^{\cl}_{xx}$ and $\Delta$, and write $\Delta$ below for the DVM-consistent, plateau-corrected scalar excess.  The reported fields therefore measure shock-layer nonequilibrium content rather than finite-grid equilibrium artefacts.

The upstream state is normalised by $\rho_1=T_1=1$ and $u_1=M_1\sqrt{\gamma T_1}$ with $\gamma=5/3$.  The downstream state is set by the monatomic Rankine--Hugoniot relations.  Coordinates are reported in upstream mean-free-path units, $(x-x_s)/\lambda_1$, where $x_s$ is the density-midpoint shock location.  The DVM references, generated with deterministic velocity-space quadrature in the spirit of discrete-velocity kinetic solvers \citep{Mieussens2000,FilbetJin2010,DimarcoPareschi2014}, use domains $[-40,40]$, $[-60,60]$ and $[-80,80]$, with $N_x=1600,1800,2200$ cells and tensor-product velocity grids $97\times19\times19$, $121\times23\times23$ and $141\times27\times27$ for $M_1=2,3,5$, respectively.  Full distributions are stored for the budget evaluations.  The spatial budget reconstructions in \S\ref{sec:heatflux}--\ref{sec:two_budget} use the BGK DVM references because they provide a clean relaxation source for both the heat-flux and scalar-excess budgets.  Collision-model robustness is then examined at the source and channel level: the Shakhov target changes the heat-flux relaxation but leaves the scalar-excess source neutral, while the ES-BGK and ES-FP targets produce stress-dependent scalar-excess corrections through, respectively, an anisotropic Gaussian covariance and anisotropic Fokker--Planck diffusion.

For the direct collisional test in \S\ref{sec:dsmc}, we use an independent one-dimensional physical, three-dimensional velocity DSMC code for a monatomic hard-sphere normal shock.  The DSMC calculation is not used to define the BGK budget; it is used to ask whether the scalar-excess production term can be measured from binary collisions themselves.  The Mach-3 case uses a domain $[-60,60]\lambda_1$ with 400 cells, an upstream target of 120 simulation particles per cell, time step $\Delta t=0.005$ in upstream mean-free-path/thermal-speed units, and Maxwellian reservoir boundaries at the Rankine--Hugoniot upstream and downstream states.  Eight independent realisations, with random seeds 1--8, are run for 180000 steps.  Sampling begins after 70000 steps, every 10 steps thereafter, giving 11001 samples per realisation and a sampling time of 550.005.  Before ensemble averaging, the profiles are aligned by the density-midpoint shock location to avoid smearing derivative-sensitive quantities such as $\partial_x J_\Delta$.  Accepted binary collisions are sampled in the active shock layer with an average count of approximately $8.2\times10^5$ per cell in the ensemble used below.

\begin{figure}
  \centerline{\includegraphics[width=0.92\textwidth,trim={0cm 0cm 0cm 1.0cm},clip]{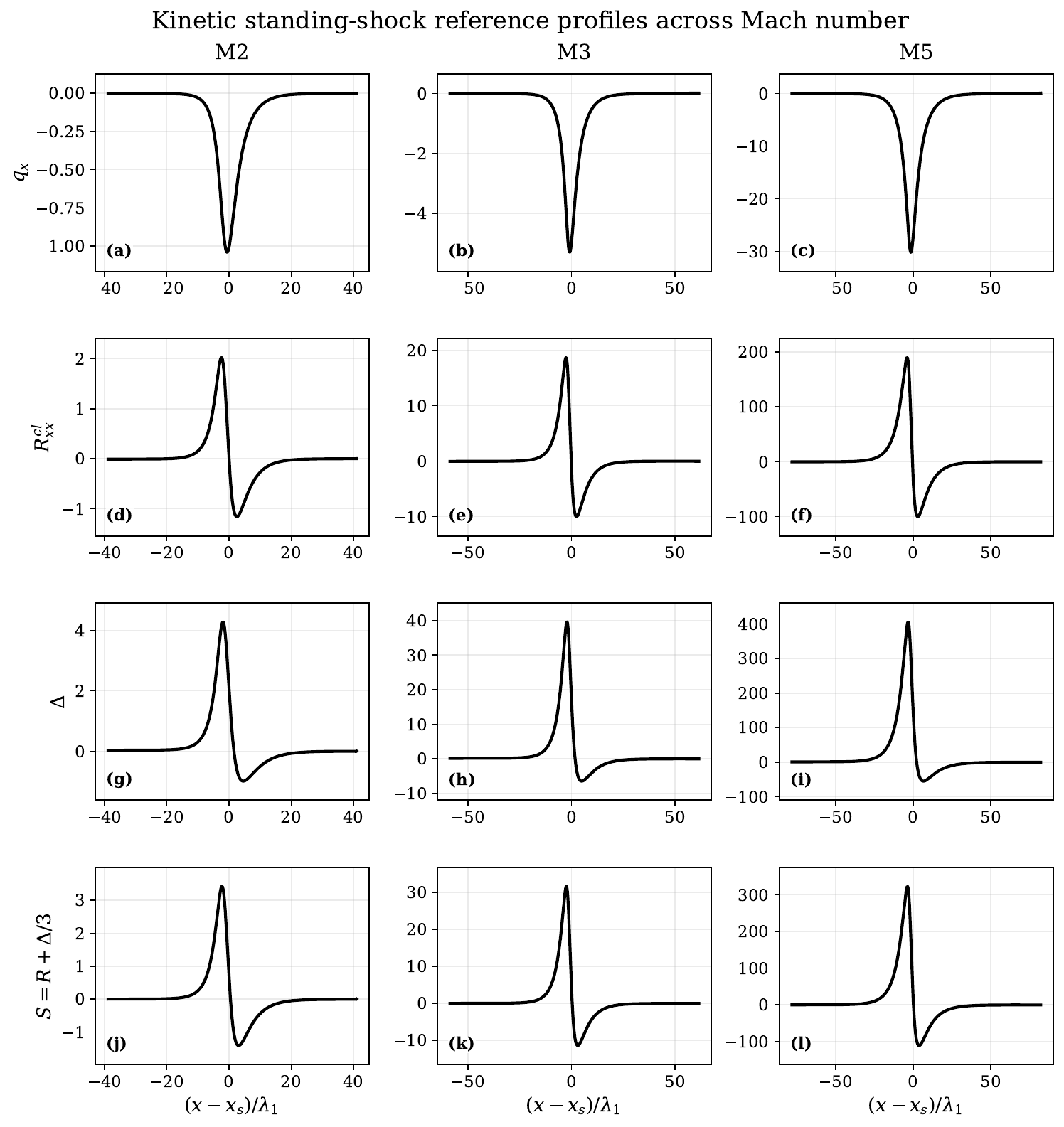}}
  \caption{Kinetic standing-shock reference profiles from conservative DVM calculations for upstream Mach numbers $M_1=2$, 3 and 5. The fields are the heat flux $q_x$, tensorial fourth-order closure moment $R^{\cl}_{xx}$, scalar fourth-order excess $\Delta$ and combined channel $S=R^{\cl}_{xx}+\Delta/3$.}
  \label{fig:dvm_ref}
\end{figure}

Figure~\ref{fig:dvm_ref} shows the fourth-order hierarchy.  The quantities $q_x$, $R^{\cl}_{xx}$, $\Delta$ and $S$ are shock-local and sign-changing, and their magnitudes grow rapidly with Mach number.  The sign changes are tail-weighted cancellations: low-order moments are nearly blind to them, whereas fourth-order kernels amplify them.  This growth is important: the ambiguity identified below is not a small visual correction to a macroscopic shock, but an order-one issue in the fourth-order closure content.

\section{Observation operator and closure-channel identifiability}
\label{sec:identifiability}

The inverse question is most transparent when separated from the numerical reconstruction.  Let the fourth-order closure state in a one-dimensional normal shock be
\begin{equation}
  \boldsymbol z(x)=\begin{pmatrix}R^{\cl}_{xx}(x)\\ \Delta(x)\end{pmatrix} .
  \label{eq:z_state}
\end{equation}
The fourth-order part of the heat-flux flux observes only
\begin{equation}
  \mathcal H\boldsymbol z = R^{\cl}_{xx}+\Delta/3 \equiv S .
  \label{eq:obs_operator}
\end{equation}
Thus $\mathcal H=[1,1/3]$ is a rank-one observation map from a two-component fourth-order state to one scalar channel.  Equivalently,
\begin{equation}
  R^{\cl}_{xx}\rightarrow R^{\cl}_{xx}+\eta,\qquad
  \Delta\rightarrow\Delta-3\eta
  \label{eq:null_transform}
\end{equation}
leaves $S$ unchanged for any shock-local field $\eta(x)$.  The null space is therefore
\begin{equation}
  \mathcal N(\mathcal H)=\operatorname{span}\left\{\begin{pmatrix}1\\ -3\end{pmatrix}\right\} .
  \label{eq:nullspace}
\end{equation}
A residual that observes only $S$ cannot distinguish two states that differ by \eqref{eq:null_transform}.  This is the precise sense in which the heat-flux budget is rank-deficient in the pair $(R^{\cl}_{xx},\Delta)$.

The same statement holds if the observed quantity is the differential heat-flux residual rather than $S$ itself.  The residual determines $\partial_x S$; after imposing the equilibrium plateau condition it determines $S$ up to the irrelevant constant fixed by the plateaux, but it still does not determine the split between $R^{\cl}_{xx}$ and $\Delta$.  Thus the difficulty is not a numerical integration constant.  It is an intrinsic non-uniqueness in the closure variables observed by the heat-flux channel.  This conclusion is also independent of whether the right-hand side is BGK, Shakhov, ES-BGK or Boltzmann.  The collision operator determines the source term used to compute the observed channel, but not the fact that the heat-flux flux contains the composite fourth-order object rather than the two closure variables separately.

A second scalar observation removes the ambiguity if and only if it is not collinear with $\mathcal H$.  Suppose an auxiliary channel observes
\begin{equation}
  \mathcal P\boldsymbol z = aR^{\cl}_{xx}+b\Delta .
  \label{eq:aux_channel}
\end{equation}
Then the pair $(\mathcal H,\mathcal P)$ is invertible exactly when
\begin{equation}
  \det\begin{pmatrix}1 & 1/3\\ a & b\end{pmatrix}= b-a/3 \neq 0 .
  \label{eq:invert_condition}
\end{equation}
The scalar-excess channel corresponds to $(a,b)=(0,1)$ and is therefore a minimal, well-conditioned complement: it supplies one scalar field and yields directly
\begin{equation}
  R^{\cl}_{xx}=S-\Delta/3 .
  \label{eq:R_from_S_Delta}
\end{equation}
This algebra explains why sparse scalar-excess probes can work: the missing information is a one-field complement to the observed heat-flux projection, not a dense direct measurement of the tensorial closure moment.

In multidimensional form, the heat-flux hierarchy contains the divergence of the composite fourth-order tensor
\begin{equation}
  A_{ij}=R^{\cl}_{ij}+\frac{1}{3}\Delta\delta_{ij} .
  \label{eq:Aij}
\end{equation}
For a one-dimensional normal shock, only $A_{xx}$ enters the streamwise heat-flux budget, reducing \eqref{eq:Aij} to \eqref{eq:obs_operator}.  In higher dimensions the observable is not the full tensor $A_{ij}$ but its divergence in the heat-flux equations; additional boundary or constitutive information is needed to recover all tensorial components.  The one-dimensional shock is therefore the cleanest setting in which the null space can be exposed and then closed by a single scalar channel.

\section{Heat-flux budget and exact recovery of the observed channel}
\label{sec:heatflux}

Multiplying \eqref{eq:bgk} by $\frac12|\boldsymbol c|^2c_x$ and integrating over velocity gives
\begin{equation}
  \partial_x F_q[\rho,u_x,T,q_x,\sigma_{xx},R^{\cl}_{xx},\Delta]
  =Q_q^{\mathrm{BGK}},
  \qquad
  Q_q^{\mathrm{BGK}}=-\frac{q_x}{\Kn} .
  \label{eq:heatbalance_general}
\end{equation}
The factor on the right follows from the same $\frac12|\boldsymbol c|^2c_x$ kernel used in the definition of $q_x$.  The fourth-order part of the flux can be written as
\begin{equation}
  F_q=u_xq_x+\frac{1}{2}B_{xx},
  \qquad
  B_{xx}=5\rho T^2+7T\sigma_{xx}+R^{\cl}_{xx}+\Delta/3 .
  \label{eq:Bxx_split}
\end{equation}
After lower-order terms are moved to the right-hand side, the fourth-order observable is precisely $S=R^{\cl}_{xx}+\Delta/3$.  The heat-flux equation therefore provides a scalar fourth-order channel, not two independent R26-level variables.

\begin{figure}
  \centerline{\includegraphics[width=0.78\textwidth]{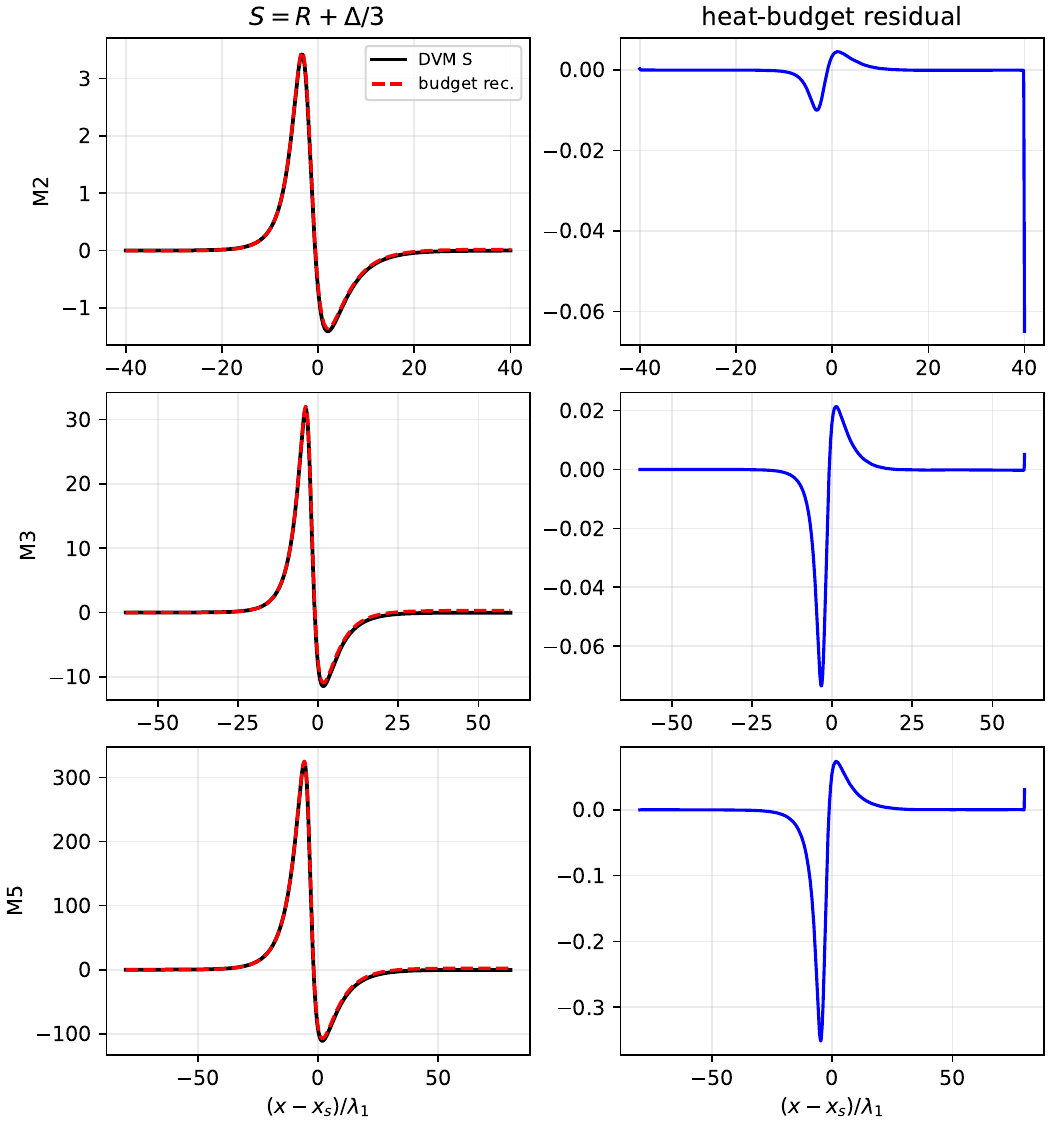}}
  \caption{Exact heat-flux-budget validation. The budget reconstructs the observed fourth-order channel $S=R^{\cl}_{xx}+\Delta/3$, not $R^{\cl}_{xx}$ alone. The residual panel verifies the budget on its own spatial-derivative scale; this figure establishes channel identification, while recovery of $R^{\cl}_{xx}$ requires the scalar-excess channel introduced in \S\ref{sec:delta_budget}.}
  \label{fig:budget}
\end{figure}

We first validate the channel before using it for closure recovery.  The exact-budget reconstruction computes the flux from the full DVM distribution and decomposes it into lower-order terms and $S$.  Figure~\ref{fig:budget} shows direct DVM $S$ and budget-reconstructed $S$ for the three Mach numbers, along with the residual.  The active zone is defined by $|S^{\DVM}|>0.02\max |S^{\DVM}|$, excluding near-equilibrium plateaux where relative errors are ill-conditioned but physically immaterial.  The budget reconstructs $S$ with active-zone errors of order a few percent, setting the accuracy scale for spatial derivative reconstructions.

\section{Scalar-excess budget and collision-model dependence}
\label{sec:delta_budget}

The missing channel is supplied by a scalar fourth-order budget.  Multiplication of \eqref{eq:bgk} by $|\boldsymbol c|^4$ gives
\begin{equation}
  \int |\boldsymbol c|^4 v_x\partial_x f\,\dd\boldsymbol v = Q_\Delta .
  \label{eq:delta_raw}
\end{equation}
Because the peculiar velocity depends on the local velocity $u_x(x)$, the flux derivative contains a moving-frame contribution.  Define
\begin{equation}
  J_\Delta=\int v_x|\boldsymbol c|^4 f\,\dd\boldsymbol v,
  \qquad
  K_\Delta=\int v_x|\boldsymbol c|^2 c_x f\,\dd\boldsymbol v .
  \label{eq:J_K_delta}
\end{equation}
Since $\partial_x|\boldsymbol c|^4=-4|\boldsymbol c|^2c_x\partial_xu_x$, one obtains
\begin{equation}
  \int |\boldsymbol c|^4 v_x\partial_x f\,\dd\boldsymbol v
  =\partial_x J_\Delta+4(\partial_xu_x)K_\Delta .
  \label{eq:delta_flux_identity}
\end{equation}
For BGK, the collision source is
\begin{equation}
  Q_\Delta^{\mathrm{BGK}}=\frac{1}{\Kn}\int |\boldsymbol c|^4(M_d-f)\,\dd\boldsymbol v=-\frac{\Delta}{\Kn} ,
  \label{eq:delta_bgk_source}
\end{equation}
where in the DVM implementation $M\to M_d$ and the identity is exact under the same $M_d$-subtraction used in \eqref{eq:Deltadef}.  Hence
\begin{equation}
  \Delta_{\rm bud}=-\Kn\left[\partial_xJ_\Delta+4(\partial_xu_x)K_\Delta\right].
  \label{eq:delta_rec}
\end{equation}
The derivative is evaluated after a short centred smoothing over seven grid points; the same choice is used for all Mach numbers.  For the Mach-3 case, increasing the smoothing window from 7 to 9, 11 and 15 points changes the active-zone $\Delta$-budget error from $3.41\%$ to $3.66\%$, $3.99\%$ and $4.89\%$, with correlations remaining above 0.9994.  The reconstruction is therefore not qualitatively tuned to a single derivative window.

The form of the scalar-excess source depends on the collision operator, but the flux-side observation operator does not.  For a generic kinetic equation $v_x\partial_x f=\mathcal C[f]$, the heat-flux source is
\begin{equation}
  Q_q=\int \frac12 |\boldsymbol c|^2 c_x\,\mathcal C[f]~\dd\boldsymbol v,
  \qquad
  Q_\Delta=\int |\boldsymbol c|^4\,\mathcal C[f]~\dd\boldsymbol v .
  \label{eq:generic_sources}
\end{equation}
The source changes with the collision model, while the observed fourth-order channel in the heat-flux flux remains the same kinematic projection.  BGK gives \eqref{eq:delta_bgk_source}.  The Shakhov target changes the heat-flux moment so that $Q_q=-Pr\,q_x/\Kn$, but its correction is odd in $\boldsymbol c$ and integrates to zero against the even kernel $|\boldsymbol c|^4$; therefore $Q_\Delta=-\Delta/\Kn$ is unchanged.  This is the analytical reason why a correct-Prandtl Shakhov check can alter the heat-flux relaxation without changing the scalar-excess source.

The ES-BGK model is different in a physically useful way.  Its target is an anisotropic Gaussian \citep{Holway1966ESBGK,FeiLiuLiuZhang2020ShockBenchmark},
\begin{equation}
  G_\Theta(\boldsymbol c)
  =\frac{\rho}{(2\pi)^{3/2}(\det\Theta)^{1/2}}
  \exp\left[-\frac12 c_i(\Theta^{-1})_{ij}c_j\right],
  \label{eq:es_gaussian}
\end{equation}
with covariance tensor $\Theta$.  In the conventional monatomic ES-BGK construction, $\Theta=T I+B$, where $B$ is a trace-free tensor proportional to the stress; in the notation used here one may write $B=\alpha_{\rm ES}\sigma/\rho$, up to the sign convention used for the stress tensor.  The heat-flux observation channel does not change, because it is determined by the left-hand-side flux.  The scalar-excess source does change, because the target distribution no longer has the Maxwellian fourth scalar moment.

For any zero-mean Gaussian with covariance $\Theta$,
\begin{equation}
  \int |\boldsymbol c|^4G_\Theta\,\dd\boldsymbol v
  =\rho\left[(\tr\Theta)^2+2\tr(\Theta^2)\right].
  \label{eq:es_fourth}
\end{equation}
This identity follows from Wick's theorem: $\langle c_i c_i c_j c_j\rangle=(\tr\Theta)^2+2\tr(\Theta^2)$.  For a Maxwellian, $\Theta=T I$, so $(\tr\Theta)^2=9T^2$ and $2\tr(\Theta^2)=6T^2$, giving $15T^2$.  For ES-BGK, however, the trace-free part $B$ gives
\begin{equation}
  \Delta_G^{\rm ES}
  =\rho\left[(\tr\Theta)^2+2\tr(\Theta^2)-15T^2\right]
  =2\rho\,\tr(B^2)
  =\frac{2\alpha_{\rm ES}^2}{\rho}\tr(\sigma^2).
  \label{eq:es_delta_target}
\end{equation}
Thus the ES-BGK scalar-excess source is not simply $-\Delta/\Kn$, but
\begin{equation}
  Q_\Delta^{\rm ES-BGK}
  =\frac{1}{\Kn}\left(\Delta_G^{\rm ES}-\Delta\right) .
  \label{eq:es_delta_source}
\end{equation}
In a one-dimensional normal shock, $\sigma_{yy}=\sigma_{zz}=-\sigma_{xx}/2$, so $\tr(\sigma^2)=3\sigma_{xx}^2/2$ and
\begin{equation}
  \Delta_G^{\rm ES}=\frac{3\alpha_{\rm ES}^2}{\rho}\sigma_{xx}^2 .
  \label{eq:es_1d_delta_target}
\end{equation}
This term is positive definite: the ES-BGK target carries a non-zero scalar fourth-order excess whenever the covariance is anisotropic.  This is the sharp distinction from Shakhov.  Shakhov changes the odd heat-flux channel and leaves the even $|\boldsymbol c|^4$ channel neutral; ES-BGK changes the even scalar-excess target through a stress-squared covariance effect.

To estimate the size of this correction in the present shocks, we post-process the Mach 2, 3 and 5 DVM profiles using \eqref{eq:es_1d_delta_target} with $Pr=2/3$ and $\alpha_{\rm ES}=1-1/Pr=-1/2$.  The resulting ES-BGK correction is reported in table~\ref{tab:model_source_corrections}.  The ratio $\|\Delta_G^{\rm ES}\|/\|\Delta\|$ in the active shock layer is approximately $4.3$--$4.5\%$ for all three Mach numbers, and the ES-BGK scalar-excess source remains almost collinear with the BGK/Shakhov source, with correlations exceeding $0.999$.  The correction is therefore modest for these monatomic shocks, but it is systematically non-zero and grows in absolute magnitude with Mach number: the peak value of $\Delta_G^{\rm ES}$ increases from $0.196$ at $M_1=2$ to $20.25$ at $M_1=5$.  This source-level diagnostic supports the analytical distinction without requiring a full ES-BGK shock solve: the observation channel is unchanged, whereas the scalar-excess production law depends on the target covariance.

Fokker--Planck models provide a complementary example because their sources are generated by drift and diffusion rather than by relaxation to a target distribution.  For a velocity-space Fokker--Planck operator
\begin{equation}
  \mathcal C_{\rm FP}[f]
  =-\partial_{c_i}(A_i f)+\frac12\partial_{c_i}\partial_{c_j}(D_{ij}f),
  \label{eq:fp_operator}
\end{equation}
the production of the fourth scalar kernel $\phi=|\boldsymbol c|^4$ is obtained by integrating the adjoint generator against $f$:
\begin{equation}
  Q_\Delta^{\rm FP}
  =\int \left[4A_i|\boldsymbol c|^2c_i+4D_{ij}c_i c_j+2D_{ii}|\boldsymbol c|^2\right]f~\dd\boldsymbol v .
  \label{eq:fp_delta_source_general}
\end{equation}
This form is useful because it makes the difference from BGK-type relaxation explicit.  In an ES-FP model with standard linear drift $A_i=-\nu c_i$ and anisotropic diffusion $D_{ij}=2\nu(T\delta_{ij}+\beta\sigma_{ij}/\rho)$, the one-dimensional shock reduction gives
\begin{equation}
  Q_\Delta^{\rm ES-FP}
  =\nu\left[-4\Delta+\frac{8\beta}{\rho}\tr(\sigma^2)\right]
  =4\nu\left[-\Delta+\frac{2\beta}{\rho}\tr(\sigma^2)\right].
  \label{eq:esfp_delta_source}
\end{equation}
For the correct-Prandtl choice $\beta=-5/4$, the ES-FP correction is therefore proportional to the same stress-squared invariant as the ES-BGK correction, but with a different coefficient and opposite sign.  This is the scalar-excess analogue of the high-order-moment production differences emphasised in shock-wave benchmark studies of BGK, Shakhov, ES-BGK, ES-FP and Cubic-FP models \citep{FeiLiuLiuZhang2020ShockBenchmark}.  It also explains why equal Prandtl number does not imply equal fourth-order closure production: the heat-flux relaxation rate is a lower-order transport constraint, whereas $Q_\Delta$ samples an even fourth-order kernel.

The constant ratio reported in table~\ref{tab:model_source_corrections} follows directly from this algebra, not from a separate numerical fit.  After the ES-FP source is scaled by $4\nu$, its stress-quadratic correction is
\begin{equation}
  Q_{\Delta,\mathrm{corr}}^{\rm ES-FP}/(4\nu)
  =\frac{2\beta}{\rho}\tr(\sigma^2),
\end{equation}
whereas the ES-BGK correction from \eqref{eq:es_delta_target} is
\begin{equation}
  \Delta_G^{\rm ES}
  =\frac{2\alpha_{\rm ES}^2}{\rho}\tr(\sigma^2).
\end{equation}
Hence their pointwise ratio is
\begin{equation}
  \frac{Q_{\Delta,\mathrm{corr}}^{\rm ES-FP}/(4\nu)}{\Delta_G^{\rm ES}}
  =\frac{\beta}{\alpha_{\rm ES}^2}.
  \label{eq:esfp_esbgk_ratio}
\end{equation}
For $Pr=2/3$, $\alpha_{\rm ES}=1-1/Pr=-1/2$ and $\beta=-5/4$, so \eqref{eq:esfp_esbgk_ratio} gives exactly $-5$.  The nearly constant $-5.0$ values in table~\ref{tab:model_source_corrections} are therefore an analytical consequence of the ES-BGK Gaussian covariance and ES-FP anisotropic diffusion coefficients; the DVM post-processing only measures the size of the common stress-squared invariant relative to the actual shock-layer $\Delta$.

\begin{figure}
  \centerline{\includegraphics[width=0.98\textwidth]{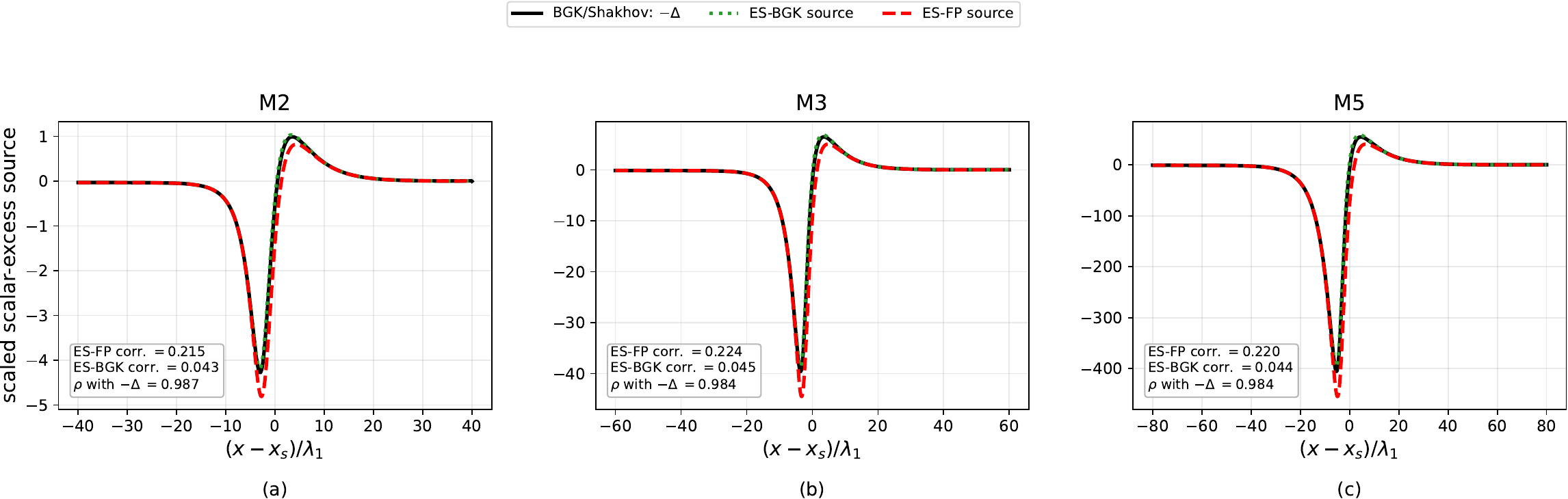}}
  \caption{Scaled scalar-excess source for BGK/Shakhov, ES-BGK and ES-FP, evaluated on the Mach 2, 3 and 5 DVM shock profiles.  The black curve is the BGK/Shakhov baseline $-\Delta$.  ES-BGK gives a small stress-quadratic correction from the anisotropic Gaussian target, whereas ES-FP gives a larger anisotropic-diffusion correction of opposite sign.  The nearly coincident shapes show that the source remains strongly aligned with the scalar-excess channel, while the separation between green and red curves shows that correct-Prandtl kinetic models can assign different fourth-order production laws to the same observed heat-flux channel.}
  \label{fig:esfp_esbgk_sources}
\end{figure}

Table~\ref{tab:model_source_corrections} and figure~\ref{fig:esfp_esbgk_sources} compare the ES-BGK and ES-FP source-level diagnostics on the same DVM shock profiles.  The ES-FP correction contributes $21.5\%$, $22.4\%$ and $22.0\%$ of $\|\Delta\|$ for Mach 2, 3 and 5, respectively, about five times the ES-BGK correction in magnitude and with the opposite sign for $\beta=-5/4$.  Nevertheless, the full scaled ES-FP source remains strongly aligned with the BGK/Shakhov scalar-excess source, with correlations $0.987$, $0.984$ and $0.984$.  Thus ES-FP does not change the kinematic observation channel; it changes the source that drives the missing scalar complement.  This distinction is central to the paper: $S$ is the channel observed by the heat-flux flux, while $Q_\Delta$ is the collision-model-specific mechanism by which the scalar-excess information is produced.

\begin{table}
  \begin{center}
  \caption{Source-level ES-BGK and ES-FP scalar-excess corrections evaluated on the DVM shock profiles.  The ratios are active-zone norms relative to $\|\Delta\|$.  The ES-FP source is the scaled source $Q_\Delta^{\rm ES-FP}/(4\nu)$ in \eqref{eq:esfp_delta_source}; the correlation compares the full ES-FP scaled source with the BGK/Shakhov baseline $-\Delta$.}
  \label{tab:model_source_corrections}
  \begin{tabular}{lrrrr}
  \toprule
  Mach & ES-BGK correction & ES-FP correction & corr$(Q_\Delta^{\rm ES-FP},-\Delta)$ & ES-FP/ES-BGK correction\\
  \midrule
  M2 & 0.0431 & 0.215 & 0.987 & $-5.0$\\
  M3 & 0.0447 & 0.224 & 0.984 & $-5.0$\\
  M5 & 0.0439 & 0.220 & 0.984 & $-5.0$\\
  \bottomrule
  \end{tabular}
  \end{center}
\end{table}

This result is consistent with benchmark studies showing that correct-Prandtl kinetic models can agree in low-order transport while differing in higher-order moments and their production terms in shock waves \citep{FeiLiuLiuZhang2020ShockBenchmark}.  In the present language, ES-BGK changes the scalar-excess source through a stress-quadratic target term, ES-FP changes it more strongly through anisotropic diffusion, and Shakhov does not change it at all.  These differences are invisible if one looks only at density, velocity, stress or heat-flux relaxation rates.  They become visible only after the scalar-excess channel has been identified as the missing complement to the heat-flux observation channel.

\subsection{Shakhov source-neutrality and channel-consistency check}
\label{sec:shakhov}

\begin{figure}
  \centerline{\includegraphics[width=1.02\textwidth]{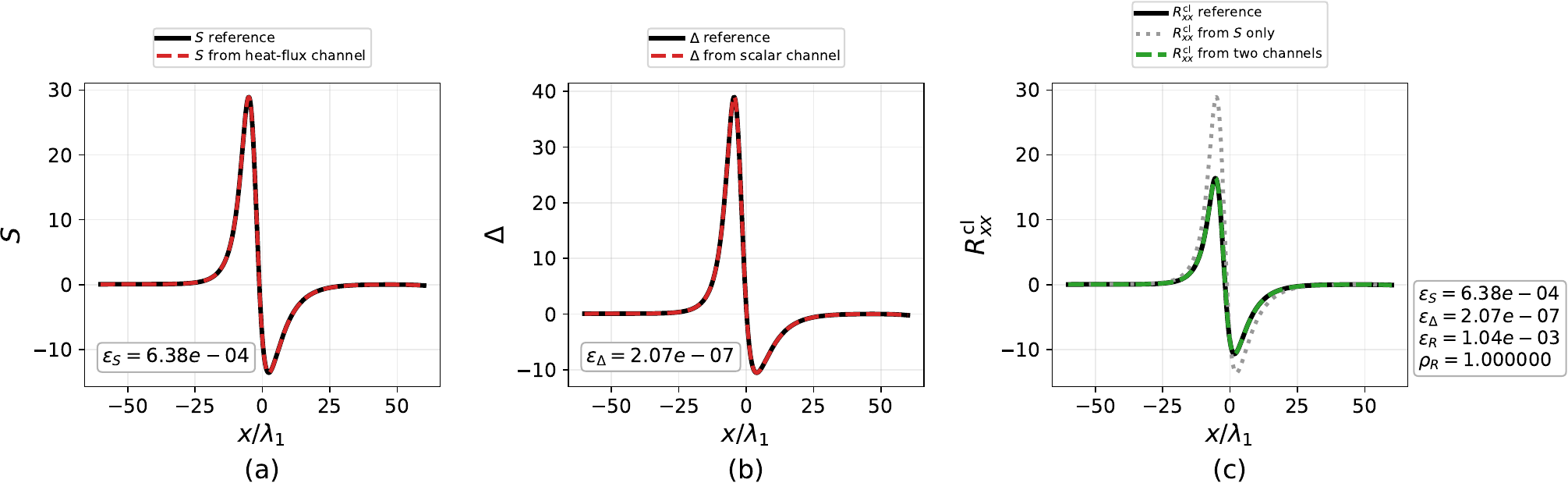}}
  \caption{Conservative discrete Shakhov channel check for the Mach-3 shock with $Pr=2/3$. (a) The heat-flux flux channel reconstructs $S$. (b) The scalar-excess source is neutral under the Shakhov correction and recovers $\Delta$. (c) The heat-flux channel alone gives a large $R^{\cl}_{xx}$ error, whereas the two-channel reconstruction recovers the tensorial closure moment. The collision sources are evaluated directly from the discrete distribution, so this checks channel consistency rather than replacing the BGK spatial budget sweep in figures~\ref{fig:two_budget} and \ref{fig:sparse_delta}.}
  \label{fig:shakhov_check}
\end{figure}

The Shakhov model is the most direct correct-Prandtl test of the BGK result because it changes the heat-flux relaxation without replacing the Maxwellian covariance by an anisotropic one.  We therefore evaluate a conservative discrete Shakhov target \citep{Shakhov1968} on the Mach-3 DVM shock.  The Shakhov target has the form of a Maxwellian multiplied by an odd heat-flux correction.  Consequently its projection onto the heat-flux kernel is non-zero, while its projection onto the even scalar kernel $|\boldsymbol c|^4$ vanishes.  Numerically, the target gives $q_S/q=0.333323\simeq 1-Pr$ and the direct heat-flux collision source gives $Q_q/q=-0.666677\simeq -Pr$, as required for $Pr=2/3$.  At the same time, the scalar-excess source remains $Q_\Delta=-\Delta/\Kn$ to numerical precision.  This check is deliberately not another spatial budget reconstruction.  It evaluates the collision projections directly from the discrete distribution and therefore isolates the channel identity from the finite-difference derivative error present in figures~\ref{fig:two_budget} and \ref{fig:sparse_delta}.  Figure~\ref{fig:shakhov_check} shows that the heat-flux flux reconstructs $S$ with a $0.064\%$ active-zone error, the scalar-excess channel reconstructs $\Delta$ to numerical precision, and the two-channel recovery reduces the $R^{\cl}_{xx}$ error from $68.4\%$ to $0.10\%$.  The point of this test is not that Shakhov is a new recovery method; it is a source-neutrality verification showing that the heat-flux observation channel and the scalar-excess complement survive correction of the Prandtl-number defect of BGK.

\subsection{Boltzmann limit and direct DSMC production}
\label{sec:model_ladder_boltzmann}

For the Boltzmann collision operator, the analogue is the genuine collision-production integral $\int |\boldsymbol c|^4 Q(f,f)\,\dd\boldsymbol v$.  In DSMC this production can be accumulated directly by summing the change of $|\boldsymbol c|^4$ over accepted binary collisions.  The hierarchy of sources is therefore: BGK and Shakhov give a scalar relaxation of $\Delta$; ES-BGK adds a stress-squared target excess; ES-FP modifies the same channel through anisotropic diffusion; and Boltzmann/DSMC supplies the true fourth-order production from molecular collisions.  In all cases, however, the flux-side heat-flux observation operator remains the same composite fourth-order channel.  The DSMC test in \S\ref{sec:dsmc} closes this loop by measuring the scalar-excess production directly in a stochastic binary-collision shock.

Table~\ref{tab:source_laws} summarises the distinction between observation and production.  The heat-flux observation channel is the same kinematic object throughout the model ladder.  The scalar-excess production law, however, changes with the collision model.  This is the organising separation used in the rest of the paper: $S$ answers what the heat-flux flux observes, while $Q_\Delta$ answers how a particular collision model supplies the missing scalar channel.

\begin{table}
  \begin{center}
  \caption{Observation channel and fourth-order source laws for the models considered here.  The heat-flux observation channel is kinematic, whereas the source laws are collision-model dependent.  Here $\psi_q=\frac12 |\boldsymbol c|^2c_x$, $G_\Theta$ is the ES-BGK anisotropic Gaussian, $A_i$ and $D_{ij}$ are Fokker--Planck drift and diffusion coefficients, and $Q(f,f)$ is the Boltzmann collision operator.}
  \label{tab:source_laws}
  \begin{tabular}{p{0.17\textwidth}p{0.22\textwidth}p{0.23\textwidth}p{0.25\textwidth}}
  \toprule
  Model & Heat-flux source $Q_q$ & Scalar-excess source $Q_\Delta$ & Interpretation\\
  \midrule
  BGK & $-q_x/\Kn$ & $-\Delta/\Kn$ & Baseline relaxation of heat flux and scalar excess.\\
  Shakhov & $-Pr\,q_x/\Kn$ & $-\Delta/\Kn$ & Corrects heat-flux relaxation; odd Shakhov correction is neutral in the even $|\boldsymbol c|^4$ channel.\\
  ES-BGK & target-dependent & $(\Delta_G^{\rm ES}-\Delta)/\Kn$ & Anisotropic covariance adds a stress-quadratic scalar-excess target.\\
  ES-FP & drift/diffusion dependent & $\nu[-4\Delta+8\beta\tr(\sigma^2)/\rho]$ & Anisotropic diffusion changes the scalar-excess production even when the Prandtl number is corrected.\\
  Boltzmann/DSMC & $\int \psi_q Q(f,f)\,\dd\boldsymbol v$ & $\int |\boldsymbol c|^4 Q(f,f)\,\dd\boldsymbol v$ & True collisional production; sampled in DSMC from pre/post-collision changes of $|\boldsymbol c|^4$.\\
  \bottomrule
  \end{tabular}
  \end{center}
\end{table}

\section{Two-budget recovery across Mach number}
\label{sec:two_budget}

\begin{figure}
  \centerline{\includegraphics[width=0.99\textwidth]{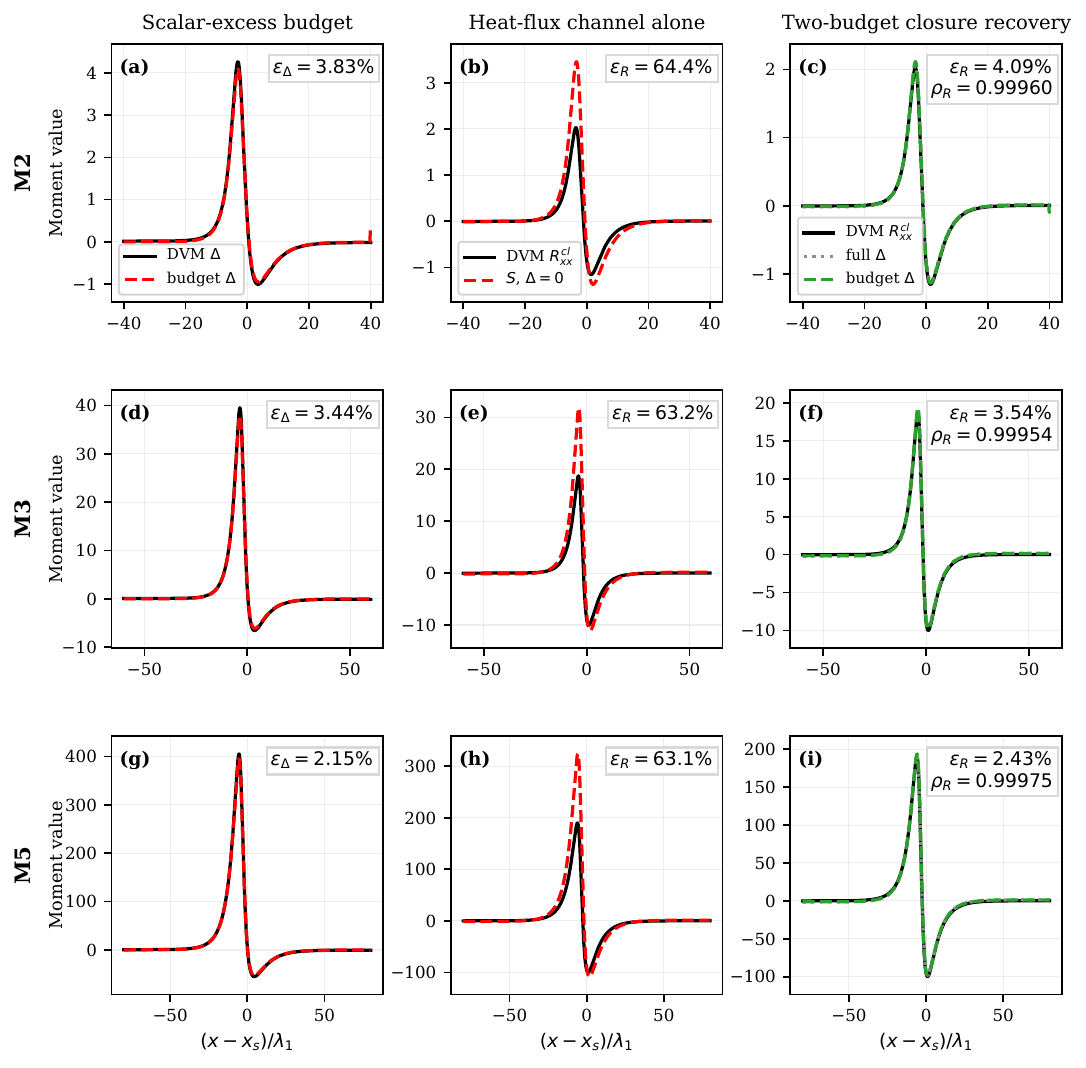}}
  \caption{Two-budget recovery of the tensorial fourth-order closure moment. First column: scalar-excess budget reconstruction of $\Delta$. Second column: heat-flux channel alone, where $R^{\cl}_{xx}\simeq S$ corresponds to setting $\Delta=0$. Third column: two-budget reconstruction $R^{\cl}_{xx}\simeq S_{\rm bud}-\Delta_{\rm bud}/3$. The dotted curve shows the full-$\Delta$ lower bound, i.e. the best reconstruction obtained when the complete DVM scalar-excess field is available.}
  \label{fig:two_budget}
\end{figure}

The closure recovery now follows directly.  The heat-flux budget gives $S$, the scalar-excess budget gives $\Delta_{\rm bud}$, and the tensorial closure is reconstructed as
\begin{equation}
  R^{\cl}_{xx,\rm bud}=S_{\rm bud}-\Delta_{\rm bud}/3 .
  \label{eq:two_budget_R}
\end{equation}
No direct $R^{\cl}_{xx}$ anchoring or supervised $R$ data enter this reconstruction.  Figure~\ref{fig:two_budget} shows the result.  The first column verifies the scalar-excess budget.  The middle column shows the rank-deficient heat-flux interpretation obtained by setting $\Delta=0$, which gives $R$ errors of about $63$--$64\%$.  The third column shows the two-budget reconstruction.  The errors drop to $4.09\%$, $3.54\%$ and $2.43\%$, close to the lower bound obtained when the full DVM $\Delta$ field is used.  The correlations exceed 0.9995.

\begin{table}
  \begin{center}
  \caption{Two-budget recovery of $R^{\cl}_{xx}$.  Errors are active-zone relative $L^2$ errors in percent.  The heat-flux budget alone gives $S$, while the scalar-excess budget supplies $\Delta$.}
  \label{tab:two_budget}
  \begin{tabular}{lrrrrr}
  \toprule
  Mach & $\epsilon_\Delta$ & $\epsilon_R(\Delta=0)$ & $\epsilon_R(\Delta_{\rm full})$ & $\epsilon_R(\Delta_{\rm bud})$ & $\rho_R(\Delta_{\rm bud})$\\
  \midrule
  M2 & 3.834 & 64.362 & 3.356 & 4.090 & 0.99960\\
  M3 & 3.435 & 63.226 & 3.102 & 3.537 & 0.99954\\
  M5 & 2.151 & 63.113 & 2.219 & 2.428 & 0.99975\\
  \bottomrule
  \end{tabular}
  \end{center}
\end{table}

\section{Sparse scalar-excess information}
\label{sec:sparse_delta}

The budget reconstruction above uses full DVM moments to establish the physical channel.  To test whether this channel represents reduced information rather than disguised direct access to $R^{\cl}_{xx}$, we replace the full $\Delta$ field by a sparse scalar-excess probe.  The DVM $\Delta$ profile is sampled at 24 shock-local points and reconstructed with a smooth radial-basis representation; no values of $R^{\cl}_{xx}$ are used.  The tensorial closure is then computed from $R^{\cl}_{xx}=S-\Delta_{\rm sparse}/3$.

\begin{figure}
  \centerline{\includegraphics[width=0.96\textwidth]{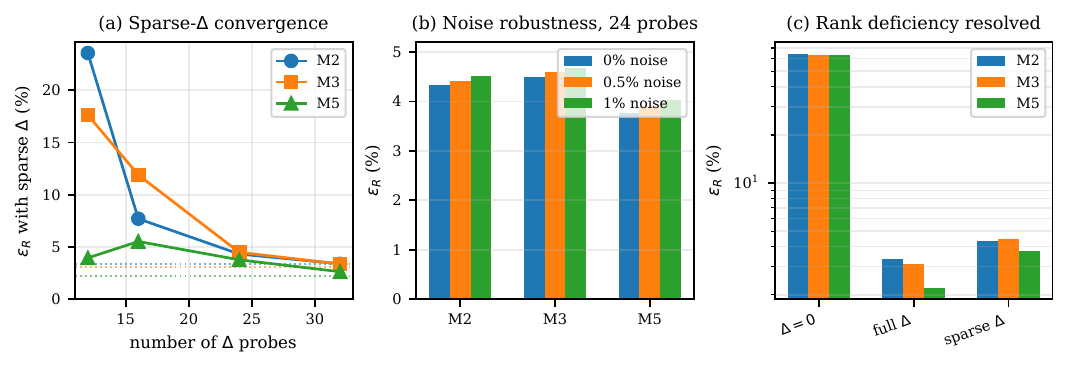}}
  \caption{Sparse scalar-excess information required to resolve the heat-flux closure ambiguity. (a) Convergence of the $R^{\cl}_{xx}$ reconstruction error as the number of shock-local $\Delta$ probes increases; dotted lines denote the full-$\Delta$ lower bounds and the 24-probe case is a representative operating point for this interpolation test in (b,c). (b) Robustness of the 24-probe reconstruction to $0.5\%$ and $1\%$ noise in the $\Delta$ probes. (c) Comparison between the rank-deficient heat-flux reconstruction $(\Delta=0)$, the full-$\Delta$ lower bound and the 24-probe sparse-$\Delta$ reconstruction.}
  \label{fig:sparse_delta}
\end{figure}

Figure~\ref{fig:sparse_delta} shows that this compact scalar channel removes most of the heat-flux rank deficiency.  With $\Delta=0$, the $R^{\cl}_{xx}$ error is $64.36\%$, $63.23\%$ and $63.11\%$ for $M_1=2,3,5$.  With 24 shock-local samples of $\Delta$ reconstructed by a smooth radial-basis interpolant, the errors fall to $4.33\%$, $4.49\%$ and $3.77\%$, close to the full-$\Delta$ lower bounds of $3.36\%$, $3.10\%$ and $2.22\%$.  Adding $1\%$ noise to the $\Delta$ probes keeps the $R^{\cl}_{xx}$ error below $4.7\%$.  This is an interpolation and information-reduction experiment, not a theorem that 24 probes are sufficient in a universal or optimal sense.  The structural statement is instead that the missing information has the type of one scalar fourth-order field; the number of probes required to interpolate that field depends on smoothness, shock thickness, noise level and probe placement.  The convergence curve in figure~\ref{fig:sparse_delta}(a) should therefore be read as an operating study for the present shock family.  Small numbers of probes already remove the order-unity ambiguity, and the 24-probe case is a practical point where the remaining error is comparable with the full-field lower bound and robust to small measurement noise.  Thus the information needed to resolve the heat-flux ambiguity is not a dense $R^{\cl}_{xx}$ field; it is an independent scalar measure of isotropic fourth-order tail intensity.  This distinction matters for experiments and reduced solvers: a scalar diagnostic, surrogate or auxiliary equation for tail intensity can close the heat-flux null space without supervising the whole tensorial closure field.

\section{Direct DSMC collision production of the scalar-excess source}
\label{sec:dsmc}

The relaxation models above are useful because they isolate specific collision-model mechanisms.  BGK and Shakhov relax the scalar excess as $-\Delta/\Kn$, while ES-BGK adds a stress-quadratic target contribution.  A stronger test is to measure the scalar-excess production from collisions without prescribing any relaxation target.  We therefore perform a standalone DSMC calculation of a Mach-3 monatomic hard-sphere normal shock and accumulate the production of the central fourth scalar moment directly during accepted binary collisions.

For each accepted collision pair in cell $j$, let $\boldsymbol c_1$ and $\boldsymbol c_2$ denote peculiar velocities relative to the instantaneous cell mean velocity before collision, and let primed quantities denote the post-collision velocities after elastic hard-sphere scattering.  The cell-wise scalar-excess production is sampled as
\begin{equation}
  Q_{\Delta,j}^{\rm DSMC}
  =\frac{1}{V_j\Delta t_{\rm samp}}
  \sum_{\rm coll\in j}
  \left[
  |\boldsymbol c'_1|^4+|\boldsymbol c'_2|^4
  -|\boldsymbol c_1|^4-|\boldsymbol c_2|^4
  \right]w_p,
  \label{eq:dsmc_qdelta}
\end{equation}
where $w_p$ is the particle weight and $\Delta t_{\rm samp}$ is the accumulated sampling time.  This quantity is the DSMC counterpart of $\int |\boldsymbol c|^4Q(f,f)\,\dd\boldsymbol v$.  It is not imposed to be proportional to $-\Delta$; it is a direct measurement of how binary collisions create or destroy the scalar-excess moment in the shock layer.

The DSMC run uses 400 spatial cells over $[-60,60]\lambda_1$, upstream particles per cell equal to 120, time step $0.005$, and eight independent seeds.  Sampling starts after 70000 steps and continues to 180000 steps, with sampling every 10 steps.  The realisations are aligned by their density-midpoint shock locations before averaging.  The resulting ensemble has 11001 samples and an average of $8.2\times10^5$ accepted collisions per active-zone cell.  This alignment is essential: the scalar-excess budget contains the derivative of the fifth-order flux $J_\Delta$, and direct averaging of unaligned shock profiles smears the derivative even when the lower-order shock structure appears converged.

\begin{figure}
  \centerline{\includegraphics[width=0.95\textwidth]{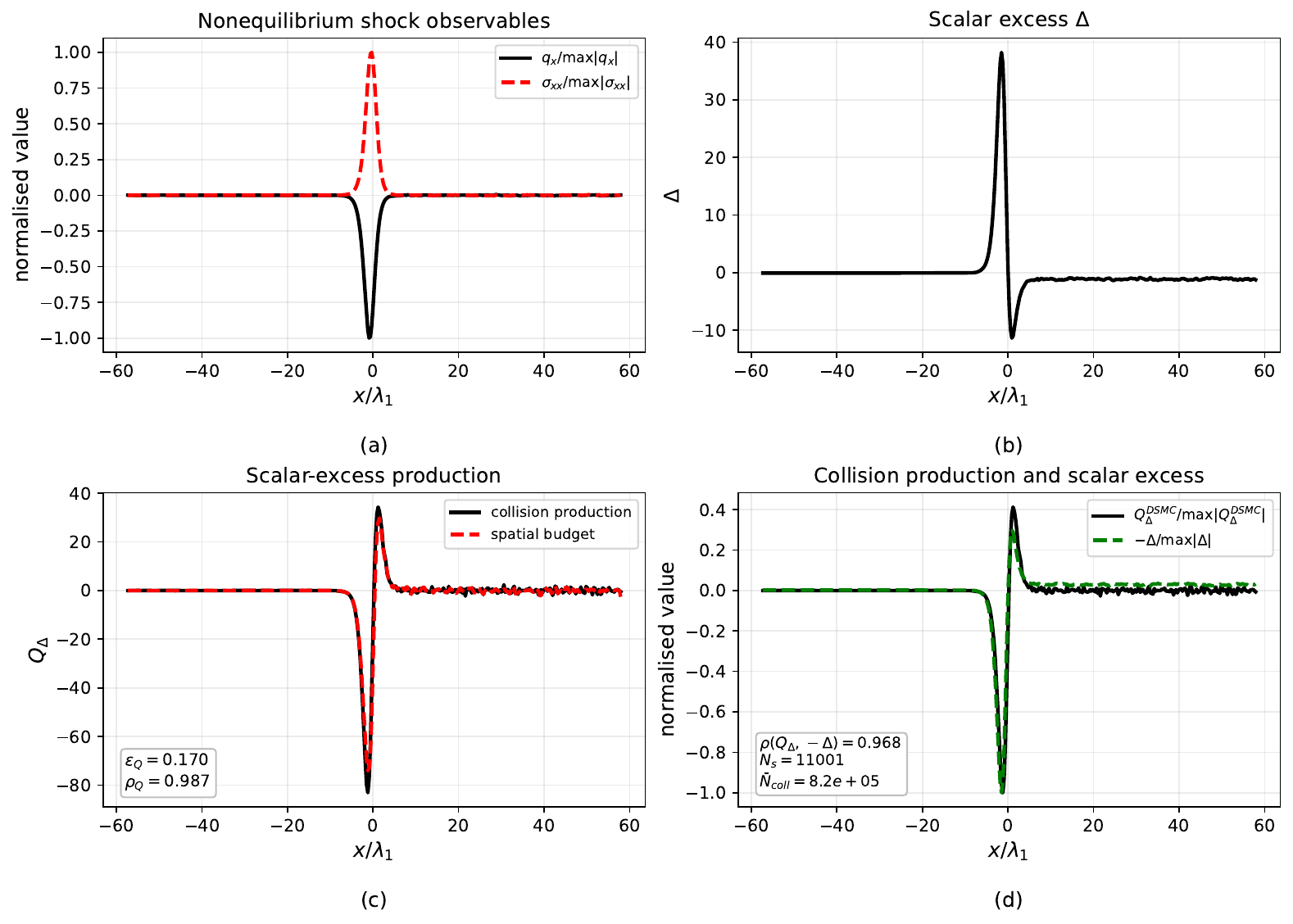}}
  \caption{Direct DSMC scalar-excess production in a Mach-3 monatomic hard-sphere shock.  (a) Normalised heat flux and normal stress identify the nonequilibrium shock layer.  (b) Scalar fourth-order excess $\Delta$.  (c) Direct collisional production $Q_\Delta^{\rm DSMC}$, accumulated from pre- and post-collision changes of $|\boldsymbol c|^4$, compared with the spatial budget $\partial_xJ_\Delta+4u_x'K_\Delta$.  The active-zone relative error is 0.170 and the correlation is 0.987.  (d) Normalised $Q_\Delta^{\rm DSMC}$ and $-\Delta$ have correlation 0.968, indicating strong relaxation-like alignment without assuming a BGK relaxation rate.}
  \label{fig:dsmc_budget}
\end{figure}

Figure~\ref{fig:dsmc_budget} shows the result.  Panel (a) verifies that the direct DSMC calculation samples the same nonequilibrium shock layer in which the heat flux and normal stress are concentrated.  Panel (b) shows the scalar excess $\Delta$.  Panel (c) compares the directly accumulated collisional production $Q_\Delta^{\rm DSMC}$ with the spatial scalar-excess budget,
\begin{equation}
  Q_\Delta^{\rm budget}=\partial_xJ_\Delta+4(\partial_xu_x)K_\Delta .
  \label{eq:dsmc_budget}
\end{equation}
The derivative is evaluated using a local cubic polynomial with a seven-point stencil; windows of 5--9 points give similar results, while larger windows over-smooth the narrow shock layer.  Specifically, stencil windows of 5, 7 and 9 give active-zone errors $0.173$, $0.170$ and $0.188$, and correlations $0.986$, $0.987$ and $0.984$, respectively.  The active-zone comparison reported in figure~\ref{fig:dsmc_budget} gives a relative error of 0.170 and a correlation of 0.987.  For DSMC, both sides of \eqref{eq:dsmc_budget} are difficult quantities: one is a noisy collision-production sample and the other contains a derivative of a fifth-order flux.  Agreement at this level is therefore a strong confirmation that the scalar-excess channel has a genuine collision-production counterpart.

Panel (d) compares the normalised collision production with the normalised $-\Delta$ profile.  The correlation is 0.968, showing that the DSMC collisions relax the scalar excess in the shock core with the same qualitative sign structure as BGK and Shakhov.  The comparison is deliberately normalised: the DSMC production is not assumed to follow the BGK/Shakhov unit relaxation rate, and the effective proportionality varies through the shock.  This is precisely the point.  The scalar channel identified by the two-budget theory is not a relaxation-model artefact; it is a collisional channel whose source law becomes model-dependent once one moves from BGK to Shakhov, ES-BGK, ES-FP and DSMC/Boltzmann physics.

The DSMC result completes the collision-model ladder used in this paper.  BGK provides the cleanest analytic budget and the Mach sweep used for recovery.  Shakhov corrects the heat-flux Prandtl relaxation while keeping the even scalar-excess source invariant.  ES-BGK changes the scalar-excess source through an anisotropic Gaussian target, and ES-FP changes it more strongly through anisotropic diffusion.  DSMC then measures the corresponding fourth-order production from binary collisions directly.  Across these models, the source law changes, but the heat-flux observation channel remains the same composite fourth-order object.  This separation between kinematic observability and collision-model production is the central point needed for closure identification.

\section{Consequences and outlook for closure modelling}
\label{sec:consequences}

The identifiability result has a direct consequence for closure modelling.  If a model is trained, fitted or regularised only through the heat-flux budget, then all states related by \eqref{eq:null_transform} are observationally equivalent to that budget.  A one-parameter family
\begin{equation}
  R^{\cl}_{xx,\eta}=R^{\cl}_{xx}+\eta(x),\qquad
  \Delta_{\eta}=\Delta-3\eta(x)
  \label{eq:closure_family}
\end{equation}
produces exactly the same observed channel $S$ for any shock-local field $\eta(x)$.  Two closures can therefore agree in the heat-flux residual and disagree in the tensorial closure moment required by an R26-type model.  This is not merely a diagnostic inconvenience.  The tensorial and scalar parts enter different members of a moment hierarchy and represent different pieces of the nonequilibrium distribution: traceless anisotropy and isotropic tail intensity.  Treating their sum as the tensorial moment itself can therefore produce a closure with the correct observed energy-transport channel and the wrong internal split.  In R26 language, such a closure would appear successful when inspected through the heat-flux equation, but it would supply the wrong tensorial fourth-order state to any equation, wall condition or regularisation term that depends on $R_{ij}$ rather than on the composite $A_{ij}=R_{ij}+\Delta\delta_{ij}/3$.

The present normal-shock results also define a compact reference case for future solvers.  A solver should first recover the observed heat-flux channel $S$; it should then supply an independent scalar-excess channel; only then should $R^{\cl}_{xx}$ be inferred.  A direct comparison of low-order fields is insufficient.  The sparse-probe experiment shows that the additional information need not be dense or tensorial, but it must be independent of the heat-flux projection.  This makes the shock problem useful as a benchmark for reduced kinetic solvers, neural closures and moment models: it tests not only whether a residual is small, but what closure content the residual observes.

For more general collision operators, the flux-side observation operator remains kinematic while the scalar-excess source changes.  The ES-BGK/ES-FP diagnostics in table~\ref{tab:model_source_corrections} and figure~\ref{fig:esfp_esbgk_sources} are useful intermediate cases: they show that correct-Prandtl models can preserve the same heat-flux observation channel while modifying the scalar-excess production law through stress anisotropy, with ES-FP producing a larger correction than ES-BGK for the same DVM profiles.  The DSMC calculation in figure~\ref{fig:dsmc_budget} is the complementary high-fidelity case: it replaces a model source by direct binary-collision production.  Together, these results show why closure identification should distinguish the observed channel from the production law.  A model may observe the same $S$ channel while assigning a different collision source to the scalar-excess complement.  In multidimensional flows, the observed heat-flux quantity is the divergence of the composite tensor $R_{ij}+\Delta\delta_{ij}/3$, so additional tensorial information is needed to reconstruct the full closure tensor.  The one-dimensional shock studied here is therefore the clean base problem, not the end point of the theory.

\section{Conclusions}
\label{sec:conclusion}

The heat-flux budget of a monatomic kinetic normal shock constrains a combined fourth-order energy-transport channel, not the tensorial closure moment alone.  In one dimension this channel is $S=R^{\cl}_{xx}+\Delta/3$, and the observation map from $(R^{\cl}_{xx},\Delta)$ to $S$ has a one-dimensional null space.  This projection property explains why residual agreement in the heat-flux equation cannot by itself certify recovery of every R26-level closure variable.

The ambiguity is constructive rather than fatal.  A DVM-consistent scalar-excess budget supplies the missing scalar field, and the two budgets recover $R^{\cl}_{xx}$ without direct $R$ data.  For BGK shocks at Mach 2--5, the error falls from about $63$--$64\%$ to $2.4$--$4.1\%$.  Sparse scalar-excess probes provide nearly the same recovery, demonstrating that the missing information is one scalar channel rather than a dense tensorial supervision signal.

The collision-model analysis shows that the observation channel and the source law must be kept separate.  Table~\ref{tab:source_laws} summarises the ladder.  Shakhov corrects the heat-flux relaxation to $Pr=2/3$ while leaving the even scalar-excess source invariant.  ES-BGK leaves the heat-flux observation channel unchanged but modifies the scalar-excess source through a stress-quadratic anisotropic-Gaussian target, contributing about $4.3$--$4.5\%$ of $\|\Delta\|$ in the present Mach 2--5 profiles.  ES-FP gives a larger anisotropic-diffusion correction of about $21.5$--$22.4\%$ while remaining strongly correlated with the BGK/Shakhov source.  Finally, DSMC measures the corresponding fourth-order production directly from accepted binary collisions.  In the Mach-3 DSMC ensemble, the direct collisional production closes the scalar-excess spatial budget with correlation 0.987 and is strongly correlated with $-\Delta$ (correlation 0.968).  Thus the scalar-excess channel is not an artefact of BGK-type relaxation targets; it is a physical collision-production channel whose specific source law changes as one moves from BGK and Shakhov to ES-BGK, ES-FP and Boltzmann/DSMC.

The result provides a closure-identifiability principle and a reference problem for testing whether kinetic closures, moment models and data-driven solvers recover the higher-order information they claim to represent.  A successful closure should not only reproduce density, temperature, stress or heat flux; it should recover the observed fourth-order channel and supply an independent scalar-excess complement before inferring the tensorial R26-level moment.

\section*{Funding}
This research received no specific grant from any funding agency, commercial or not-for-profit sectors.

\section*{Declaration of AI-assisted tools}
ChatGPT (GPT-5.5 Thinking, OpenAI) was used for language editing and for assistance in debugging numerical scripts.  The author verified all numerical results, figures, equations and scientific conclusions.

\section*{Declaration of interests}
The author reports no conflict of interest.

\section*{Data availability statement}
The DVM profiles, Shakhov channel-check data, ES-BGK and ES-FP scalar-source diagnostics, DSMC ensemble profiles, direct collision-production diagnostics, budget-reconstruction scripts and plotting scripts used to generate the figures and tables are included with the submission files for review.  A versioned public repository with a DOI will be deposited upon acceptance.

% Bibliography is included explicitly for arXiv compatibility.

\end{document}